\documentclass[twocolumn,longbib]{aastex701}
\usepackage{amsmath}
\usepackage{amssymb}
\usepackage{mathtools} 
\usepackage{graphicx} 
\usepackage[capitalize]{cleveref}
\usepackage{lineno}
\usepackage{booktabs} 
\usepackage[table]{xcolor}
\makeatletter
\g@addto@macro\bfseries{\boldmath}
\makeatother

\newcommand{\ns}{n_\mathrm{sat}}

\newcommand{\pl}{p_\mathrm{term}}
\newcommand{\nl}{n_\mathrm{term}}
\newcommand{\mul}{\mu_\mathrm{term}}

\newcommand{\ph}{p_\mathrm{pQCD}}
\newcommand{\nh}{n_\mathrm{pQCD}}
\newcommand{\muh}{\mu_\mathrm{pQCD}}

\newcommand{\pmin}{p_\text{min}}
\newcommand{\pmax}{p_\text{max}}
\newcommand{\ilh}{\mathcal{I}_\mathrm{pQCD}}

\renewcommand{\thefootnote}{\fnsymbol{footnote}}
\begin{document}

\title{As above, so below: assessing extremeness of the neutron-star equation of state \\ based on the unstable branch}
\date{\today}

\author[0000-0003-3469-7574]{Tyler Gorda}
\affiliation{Center for Cosmology and AstroParticle Physics (CCAPP), Ohio State University, Columbus, OH 43210}
\affiliation{Department of Physics, The Ohio State University, Columbus, OH 43210, USA}
\email{gorda.1@osu.edu}

\author[0000-0002-2188-3549]{Oleg Komoltsev}
\affiliation{Institut für Theoretische Physik, Goethe Universität,
Max-von-Laue-Str. 1, 60438 Frankfurt am Main, Germany}
\email{komoltsev@itp.uni-frankfurt.de}

\author[0000-0001-7991-3096]{Aleksi Kurkela}
\affiliation{Faculty of Science and Technology, University of Stavanger, 4036 Stavanger, Norway}
\email{aleksi.kurkela@uis.no}

\author[0000-0002-0079-6841]{Jürgen Schaffner-Bielich}
\affiliation{Institut für Theoretische Physik, Goethe Universität,
Max-von-Laue-Str. 1, 60438 Frankfurt am Main, Germany}
\email{schaffner@astro.uni-frankfurt.de}

\begin{abstract}
    Microscopic models of neutron-star matter have been widely used in astrophysical applications. The focus of attention has been on densities up to the maximal densities reached in stable neutron stars. The possibility that the underlying model assumptions may have important implications at higher densities has not been addressed. Here, we show that the behaviour at higher densities is strongly constrained by requiring a causal, stable, and thermodynamically consistent extension to the perturbative-QCD regime. We explicitly reveal what that behaviour must be and provide a tool for constructing and visualizing such extensions. We find that purely hadronic models trusted up to the maximal central density often require radically different behaviour at higher densities from that assumed in the original model, while models with additional degrees of freedom fare better. Our analysis disfavors purely nucleonic models for describing all stable neutron stars and supports the appearance of some type of additional degrees of freedom in stable massive neutron stars. 



\end{abstract}


\section{Introduction}
\label{sec:intro}

Neutron stars (NSs) are among the most compact astrophysical objects known and, to date, are the only objects that provide empirical access to the behavior of cold, extremely dense matter. 
Observational advances may make it possible to address the composition of matter at the densities found in the cores of these stars~\citep{Annala:2019puf,Han:2022rug,Annala:2023cwx}. 
In order to do so, the observations must be analyzed in the context of microphysical models. 
The microphysical assumptions underlying these models affect the predicted equation of state (EoS) and, ultimately, the macroscopic and measurable properties of NSs thereby allowing microphysical questions to be addressed through observations of astronomical objects.

As the EoS cannot currently be computed directly from the fundamental theory of quantum chromodynamics (QCD) at NS densities~\citep{deForcrand:2009zkb,Nagata:2021ugx}, a large number of microphysical models have been developed to facilitate such studies, some of which are available through publicly accessible databases~\citep{Typel:2013rza,Oertel:2016bki,CompOSECoreTeam:2022ddl,Antonopoulou:2022yot,MUSES_web,MUSES:2023hyz}.
These phenomenological models encode expectations based on specific physical assumptions rather than first-principles calculations with well-defined uncertainties, and are therefore typically valid only within a limited density range where they provide a meaningful description of matter.
Ideally, this range should include all densities realised in NSs (typically a few times saturation density $\ns \approx 0.16~\mathrm{fm}^{-3}$).

As we show here, microphysical models also constrain the EoS beyond the density domains in which they can be directly applied.
This somewhat unintuitive feature arises from the requirement that the EoS at NS densities be consistent with first-principles perturbative QCD (pQCD) calculations at asymptotically high densities.

For most physical quantities, pQCD does not provide robust information at intermediate or lower densities.
The EoS, however, is a special case: mechanical stability, causality, and thermodynamic consistency restrict how it can interpolate between the densities described by the models and the high-density regime constrained by pQCD~\citep{Komoltsev:2021jzg}.
As a result, these requirements impose constraints even in regions where neither the phenomenological models nor pQCD calculations can be directly applied.

We explore the behaviour that different NS-matter models \emph{predict} at densities above those reached in NSs by employing a recent non-parametric, model-agnostic prior that samples the space of EoSs connecting the model predictions at NS densities with the pQCD EoS at high densities~\citep{Gorda:2025aiu}.
By constructing a large number of interpolating EoSs, we investigate the implications of the underlying model assumptions at higher densities.
In particular, we find that many models, when extended to the maximal densities realised in NSs, imply a relatively specific continuation of the EoS at higher densities.
This continuation typically involves a strong phase-transition-like change in the behaviour of matter just above the highest densities reached in NSs.
We argue that this information can be used to further constrain EoS models.

The paper is organized as follows.
In Section~\ref{sec:sampling}, we discuss our method for sampling extensions of microphysical models up to the regime where pQCD calculations are converged. 
In Section~\ref{sec:selection}, we discuss our selection of representative microphysical models to use in this work.
In Section~\ref{sec:results}, we examine the EoS extensions and discuss what different microphysical models imply for thermodynamic behaviour on the unstable branch. 

\section{Methods}
\label{sec:method}

The NS-matter EoS, $p(\mu)$ with $p$ the pressure and $\mu$ the baryon chemical potential, can be computed in pQCD at high (baryon number) densities of around $\nh \sim 20-40\, \ns$~\citep{Gorda:2021znl,Gorda:2023usm,Gorda:2023mkk}. 
While no NS reaches these densities, the requirement to reach these results at high densities constrains how the EoS can behave at lower densities. 

In particular, given a NS-matter model extending up to some termination density, $\nl$, the requirement to reach the pQCD constraint in a causal, stable and thermodynamically consistent manner limits how the EoS can behave at intermediate densities ${\nl < n < \nh}$. 
The origin of these pQCD constraints was demonstrated in~\cite{Komoltsev:2021jzg}. 
By sampling possible interpolations between the endpoint of the NS-matter model and the pQCD limit, we can reveal what kind of behaviour is required to connect these two regimes. 

To characterize how extreme the connection between the two limits is, a useful quantity is the pQCD tension index, defined in~\cite{Komoltsev:2023zor} as
\begin{equation}
\label{eq:iqcd}
    \ilh \equiv \frac{\ph-\pl-\Delta\pmin}{\Delta\pmax-\Delta\pmin}.
\end{equation}
$\ilh$ quantifies how close the bounds for all valid EoSs connecting the low- and high-density points are to the minimal ($\ilh = 0$, with $\Delta \pmin$) or maximal ($\ilh = 1$, with $\Delta \pmax$) pressure-difference constructions between $\mul$ and $\muh$. 
These minimal and maximal constructions combine maximally causal segments with large first-order phase transitions. 
Explicit formulas for $\Delta \pmin$ and $\Delta \pmax$ are provided in~\cite{Komoltsev:2021jzg}.

\begin{figure*}[ht]
    \centering
    \includegraphics[width=1.0\linewidth]{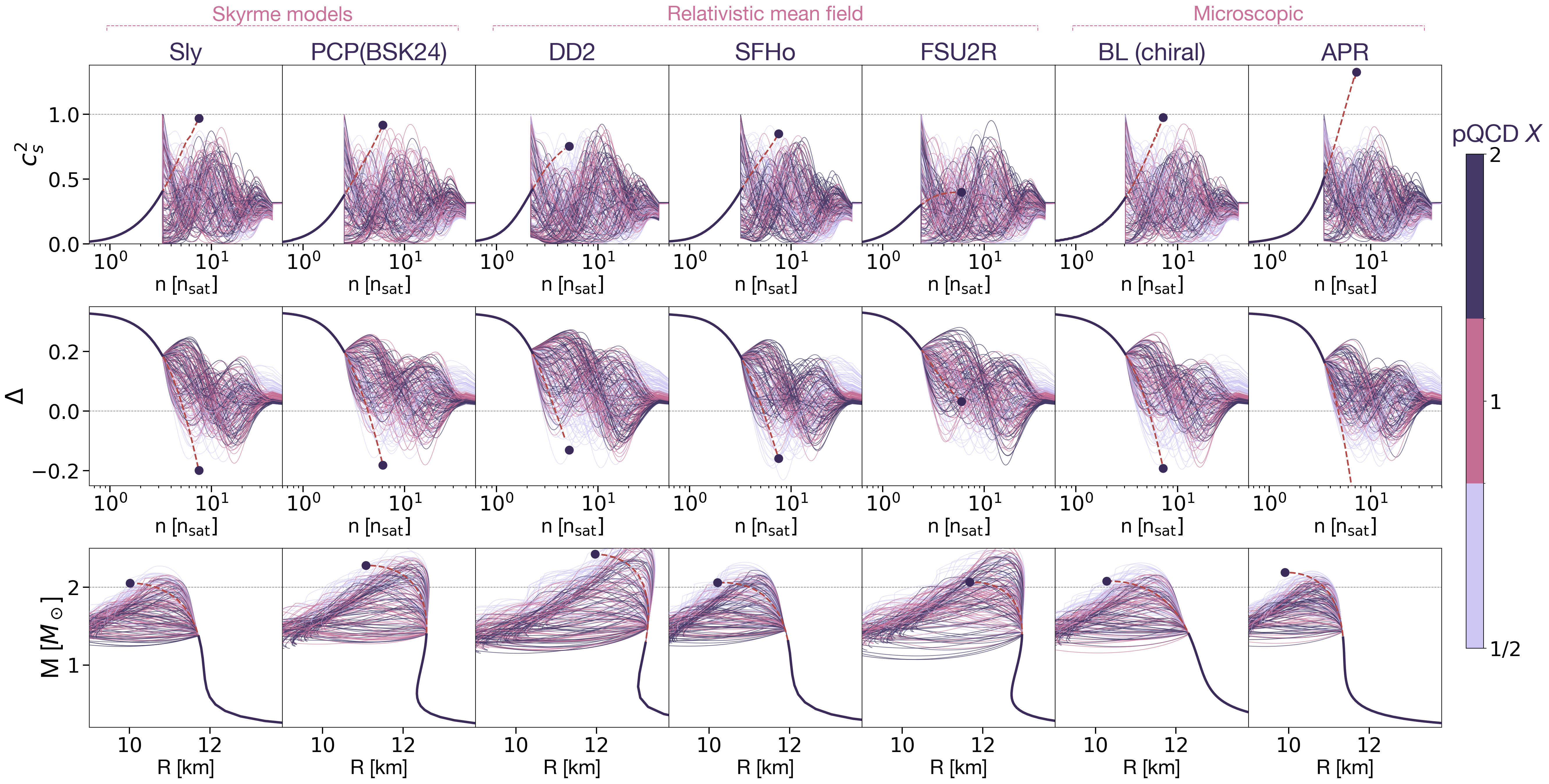}
    \caption{Possible extensions of the representative hadronic models for $c_s^2$ vs $n$, $\Delta$ vs $n$, and the mass–radius ($M-R$) relation from $1.4M_\odot$. The colors represent the value of the pQCD renormalization-scale parameter $X$. There are consistent interpolations for all values of  $X$. The dot indicates the maximal central (TOV) density.}
    \label{fig:matrix_models_14}
\end{figure*}

\subsection{Sampling of the EoS extensions}
\label{sec:sampling}

For any allowed model with $\ilh \in (0,1)$, one can define a class of EoS extensions beyond the termination density. 

Here, we construct such extensions using a Gaussian-process bridge (GPB) introduced in~\cite{Gorda:2025aiu}. 
The method first samples the allowed functional space of EoS extensions in the interval ${\nl < n < \nh}$ via a hierarchical self-similar refinement procedure, successively adding EoS points that remain consistent with all previously sampled points. 
The resulting EoSs span the allowed functional space between the low- and high-density limits and contain structures on all scales. 
The EoSs are then processed by diffusive filtering to impose a chosen correlation length for $\mu(n)$. 
By construction, this procedure efficiently samples the full space of EoS extensions of a given low-density NS EoS that are thermodynamically consistent with the high-density pQCD constraints.

The uncertainty of the pQCD results are quantified by a choice of dimensionless renormalization scale ${X \equiv 3\bar\Lambda/ (2\mu)}$ with $\bar\Lambda$ the renormalization scale in the modified minimal subtraction scheme.
The value of $X$ is usually chosen to minimize large logarithms in the perturbative results.
We use for the pQCD EoS the scale-averaged result of~\cite{Gorda:2022jvk}, taking a log-uniform distribution of $X$ values in the range $X \in [1/2 , 2]$~\citep{Gorda:2023usm}.
We note that the apparent convergence of the pQCD result is slowest for the smallest values of $X \approx 1/2$, rendering the small values less trustworthy.
Extended discussions of the convergence of the pQCD results can be found in~\cite{Gorda:2023usm,Semposki:2024vnp,Semposki:2025etb}.

In practice, we generate the self-similar EoSs extensions between the chosen termination density of the model $\nl$, and the high-density limit at ${\nh = 30\, \ns}$, where we match to the pQCD results. 
As prescribed in~\cite{Gorda:2025aiu}, we diffuse the self-similar extensions between the termination density and $40\,\ns > \nh$. 
Extending the upper diffusion region into the pQCD EoS range (from 30 to $40\,\ns$) enables a smooth variation of the sound speed when transitioning to the high-density regime. 
However, unlike in~\cite{Gorda:2025aiu}, we do not extend the diffusion into the model region at lower densities, allowing for sharp features in the EoS when transitioning from the model to the extensions. 
We consider this approach more conservative, as discussed later. 
For the final diffused extensions, we employ a hierarchical model of different logarithmically constant correlation lengths $\sigma/n$, selecting them from within a uniform range $\sigma/n \in [0.2, 0.4]$.

\begin{figure*}
    \centering
    \includegraphics[width=1.0\linewidth]{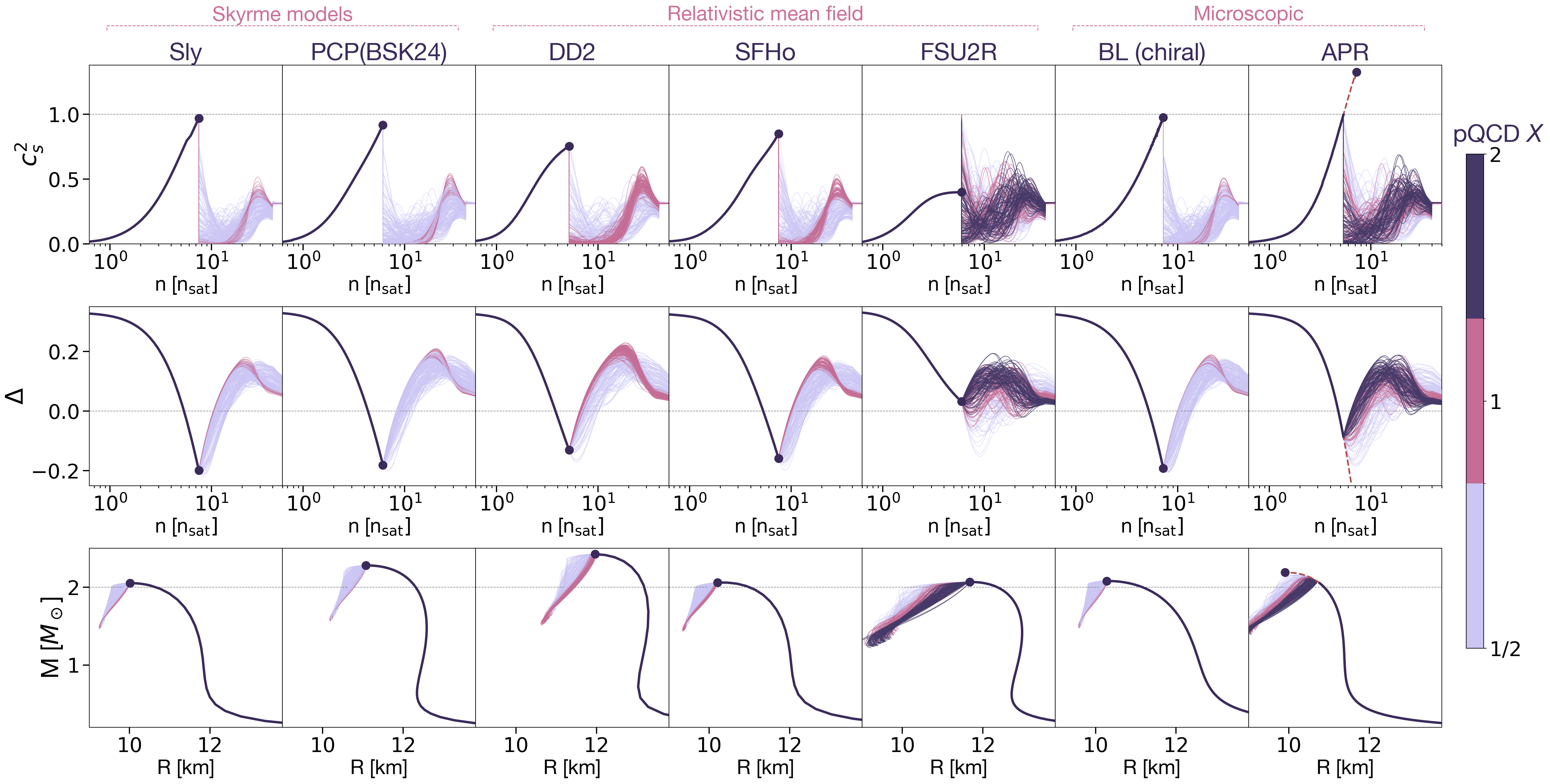}
    \caption{Possible extensions of the representative hadronic models for $c_s^2$ vs $n$, $\Delta$ vs $n$, and the mass–radius ($M-R$) relation from $M_\mathrm{TOV}$ (except for APR, which is connected from the last causal point at $M\approx2.07M_\odot$). The colors represent the value of the pQCD renormalization-scale parameter $X$. All models but FSU2R and APR can be connected only to lower values of $X$. Even in those cases, the change in behaviour is dramatic. The dot indicates the maximal (TOV) central density.}
    \label{fig:matrix_models_tov}
\end{figure*}

\subsection{Model selection and representative models}
\label{sec:selection}

We choose representative EoSs from the three classes of nucleonic models for the cold NS EoSs listed in the CompOSE database~\citep{Typel:2013rza}, namely microscopic calculations, non-relativistic density functional models, and relativistic density functional models. 
We focus on EoSs based on nuclear models fitted directly to experimental observables, i.e., nucleon-nucleon scattering data and properties of nuclei, binding energies, radii, and surface thickness.
This excludes nuclear model parameterizations based on nuclear matter properties. 
As additional criteria, we check the EoSs for consistency with the well established EoS up to saturation density and that the EoS gives a maximum NS mass of at least $2M_\odot$. In some cases, the parameter fits include the EoS of pure neutron matter. For each model, the corresponding reference and CompOSE entry are listed in \cref{table:1}.

\begin{table}[b]
\centering
\vspace{3mm}
\begin{tabular}{@{}l@{}}
\toprule
\makebox[0.20\columnwidth][l]{Name}
\makebox[0.60\columnwidth][c]{Citation}
\makebox[0.15\columnwidth][r]{CompOSE ID}\\
\midrule

\makebox[0.20\columnwidth][l]{Sly4}
\makebox[0.60\columnwidth][c]{\cite{Chabanat:1997un}}
\makebox[0.15\columnwidth][r]{134}\\

\makebox[0.20\columnwidth][l]{PCP(BSK24)}
\makebox[0.60\columnwidth][c]{\cite{Pearson:2018tkr}}
\makebox[0.15\columnwidth][r]{253}\\

\makebox[0.20\columnwidth][l]{DD2}
\makebox[0.60\columnwidth][c]{\cite{Typel:2009sy}}
\makebox[0.15\columnwidth][r]{18}\\

\makebox[0.20\columnwidth][l]{SFHo}
\makebox[0.60\columnwidth][c]{\cite{Steiner:2012rk}}
\makebox[0.15\columnwidth][r]{34}\\

\makebox[0.20\columnwidth][l]{FSU2R}
\makebox[0.60\columnwidth][c]{\cite{Negreiros_2018}}
\makebox[0.15\columnwidth][r]{214}\\

\makebox[0.25\columnwidth][l]{BL (chiral)}
\makebox[0.60\columnwidth][c]{\cite{Bombaci:2018ksa}}
\makebox[0.1\columnwidth][r]{121}\\

\makebox[0.20\columnwidth][l]{APR}
\makebox[0.60\columnwidth][c]{\cite{Akmal:1998cf}}
\makebox[0.15\columnwidth][r]{68}\\

\makebox[0.20\columnwidth][l]{BFH(QHC19-B)}
\makebox[0.60\columnwidth][c]{\cite{Baym_2019}}
\makebox[0.15\columnwidth][r]{140}\\

\makebox[0.30\columnwidth][l]{OPGR(DDHdeltaY4)}
\makebox[0.60\columnwidth][c]{\cite{Oertel_2015}}
\makebox[0.05\columnwidth][r]{67}\\

\makebox[0.25\columnwidth][l]{DS(CMF)-7}
\makebox[0.60\columnwidth][c]{\cite{Clevinger:2022xzl}}
\makebox[0.1\columnwidth][r]{194}\\

\makebox[0.20\columnwidth][l]{DD2-VQCD}
\makebox[0.60\columnwidth][c]{\cite{PhysRevX.12.041012}}
\makebox[0.15\columnwidth][r]{289}\\

\makebox[0.20\columnwidth][l]{Quarkyonic}
\makebox[0.60\columnwidth][c]{\cite{PhysRevD.102.023021}}
\makebox[0.15\columnwidth][r]{--}\\
\bottomrule
\end{tabular}
\caption{
\label{table:1}
Cold neutron-star EoS models used in \cref{fig:matrix_models_tov,fig:matrix_models_14,fig:exotica}. The corresponding CompOSE entries can be accessed at \texttt{compose.obspm.fr/eos/ID}, where \texttt{ID} denotes the CompOSE ID listed in the last column.}
\end{table}

For microscopic calculations we choose the EoSs denoted as APR~\citep{Akmal:1998cf} and BL(chiral)~\citep{Bombaci:2018ksa}.
The APR EoS is constructed from the Argonne nucleon-nucleon potential fitted to nucleon-nucleon scattering data and three-body forces plus further corrections for the calculation of the EoS. 
As a representative for an EoS built from chiral nuclear forces, again constrained by two-body and three-body nuclear data,
we take the EOS BL(chiral).

The non-relativistic density functionals, or Skyrme-type models, adopted below are the parameter set BSk24~\citep{Pearson:2018tkr} and Sly4~\citep{Chabanat:1997un,Gulminelli:2015csa}. 
The EoS BSk24 derives from Hartree--Fock--Bogoliubov nuclear mass model fitted to the binding energy of nuclei over nearly the entire chart of nuclides and the EoS of pure neutron matter. 
The Skyrme model SLy4 has been fitted to properties of selected nuclei and the EoS of pure neutron matter. 

Similar to the non-relativistic density functionals, our chosen relativistic density functional models DD2~\citep{Typel:2009sy,HEMPEL2010210} and FSU2R~\citep{Negreiros_2018} have been fitted to properties of nuclei. 
In addition, the parameters of the set FSU2R are chosen to give a reasonable description of pure neutron matter by fixing the slope parameter $L$ and have been modified from the original FSU2 model~\citep{PhysRevC.90.044305} to arrive at a smaller radius for a $1.4M_\odot$ NS,  in accord with the constraint on the tidal deformability from NS merger gravitational wave event GW170817.
The nuclear model DD2 includes density-dependent coupling constants which are tuned to describe the nonlinear density dependence of the  nucleon self-energy extracted from relativistic Brueckner--Hartree--Fock calculations. 
By this additional input, the low density pure neutron matter EoS up to saturation density can be described.

In addition to these nuclear models we investigate the effect of possible exotic phases and modern, nonconventional approaches for the cold NS EoS. 
In particular, for the EoS with exotic matter we take QHC18~\citep{Baym_2019} as a representative of an EoS with quark matter in the core, DDHdeltaY4~\citep{Oertel_2015} as an EoS with hyperonic matter appearing at high density, and  CMF-7~\citep{PhysRevC.81.045201,PhysRevC.103.025808,Clevinger:2022xzl} as a chiral mean-field model which includes hyperons and Delta-baryons as quasi-particle degrees of freedom. 
A quarkyonic model was also implemented using the publicly available code from \cite{zhao_quarkyonic_code}, based on the model described in \cite{PhysRevD.102.023021}, with the following parameters: symmetry-energy slope $L = 50\,\mathrm{MeV}$, shell parameter $\Lambda = 1400\,\mathrm{MeV}$, and quark drip density $n_t = 0.3\,\mathrm{fm}^{-3}$. 
Finally, we take an EoS using a holographic approach to describe the NS-matter EoS, namely the V-QCD EoS~\citep{PhysRevX.12.041012}.

All EoSs used in this study correspond to zero-temperature EoSs (or the lowest-temperature slice available) in $\beta-$equilibrium. 
For general-purpose EoSs, the $\beta-$equilibrium composition was recalculated according to the CompOSE prescription.

\begin{figure*}
    \centering
    \includegraphics[width=0.8\linewidth]{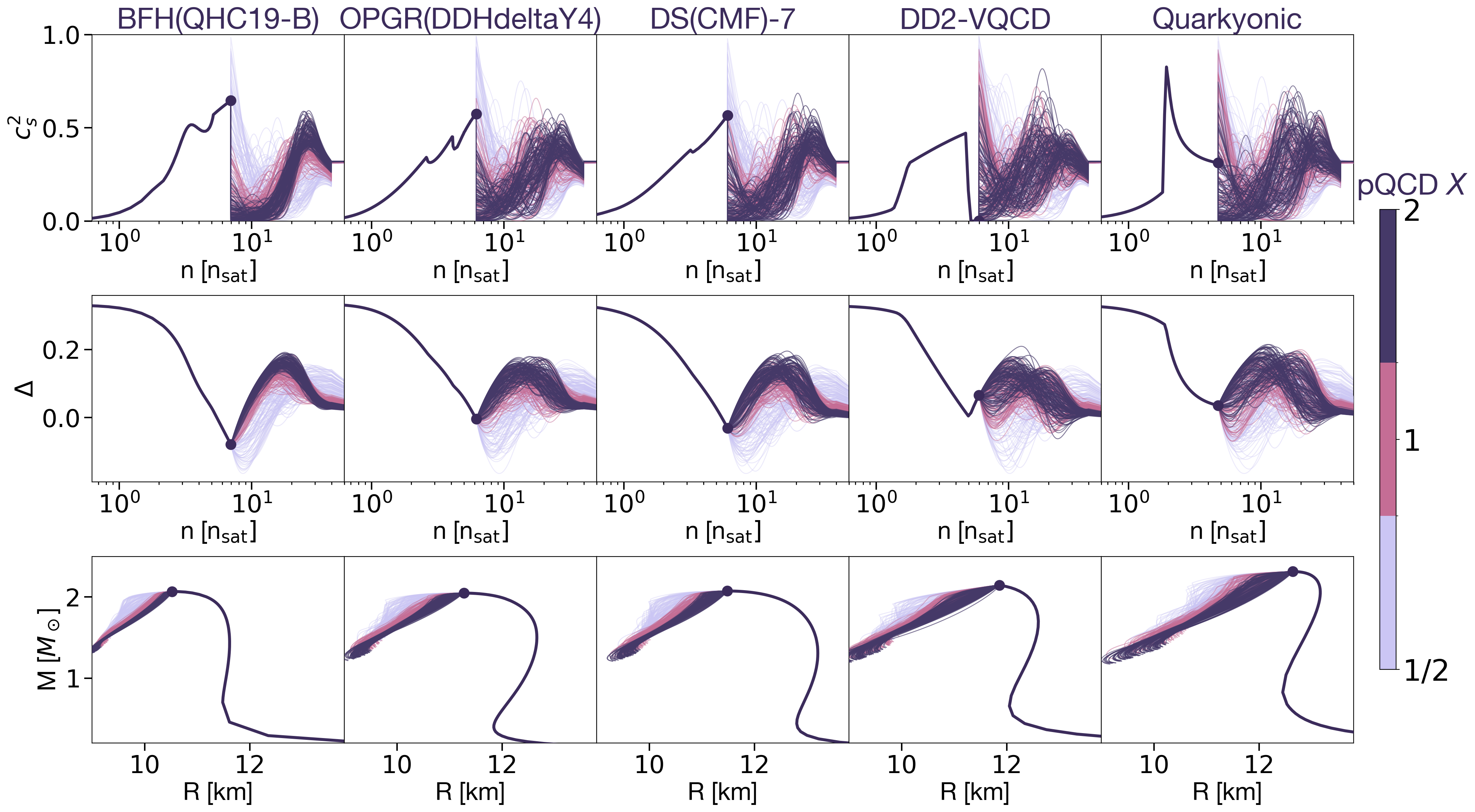}
    \caption{Possible extensions of the models featuring possible exotic phases for $c_s^2$ vs $n$, $\Delta$ vs $n$, and the mass–radius ($M-R$) relation beyond the maximal TOV density. The colors represent the value of the pQCD renormalization-scale parameter $X$. The dot indicates the TOV central density. In the absence of softening before the TOV point, the required behaviour above it becomes dramatic.}
    \label{fig:exotica}
\end{figure*}
\section{Results}
\label{sec:results}

\cref{fig:matrix_models_14} shows the possible extensions of the fully hadronic EoSs, assuming that the models remain valid up to the densities reached in a canonical $1.4M_\odot$ NS. 
The EoSs are displayed in terms of the squared speed of sound, $c_s^2$, and the (normalized) trace anomaly, ${\Delta \equiv (\epsilon - 3p)/(3\epsilon)}$, as functions of density. 
Both quantities characterize the properties of matter. 
The trace anomaly, $\Delta$, has the advantage of being obtained by integrating $n(\mu)$ and is therefore insensitive to local variations in $n(\mu)$. 
By contrast, there is no thermodynamic constraint on how rapidly $c_s^2(n)$ may vary. 

For each model, the black solid line extends up to the central density of a $1.4M_\odot$ star. 
The continuation of each model to higher densities is shown by the red dashed line, terminating at a black point that represents the core of the maximally massive star. 
At high densities, ${n = 40\, \ns}$, the EoSs are given by the pQCD calculation, corresponding to $c_s^2 \approx 1/3$ and a small $\Delta \approx 0$. 
The lines are colored based on the value of the dimensionless pQCD renormalization scale $X$ with three ranges equally spaced in $\log(X)$, i.e.,  ${X_{\rm low} \in [0.5, 0.79]}$ (light purple),  ${X_{\rm mid} \in [0.79, 1.26]}$ (magenta), and ${X_{\rm high} \in [1.26,2]}$ (dark violet).
We note that the $X_{\rm mid}$ range approximately corresponds to the `most-consistent' prediction for the next-to-next-to-next-to-leading order pQCD EoS as presented in~\cite{Gorda:2023mkk} and may be thought of as a proxy for the potential full pQCD result at this order.

The range between $n(1.4 M_\odot)$ and $40\,\ns$ is interpolated using a number of draws from the GPB described above. 
The transition to the interpolating GPB is---or at least can be---smooth, and a broad range of different possible intermediate behaviors are sampled. 
We see that there are consistent interpolations for all values of $X$. Similarly, the resulting mass-radius curves show broad range of behaviors. 

Extending these purely nucleonic models to densities reached in maximally massive stars (the TOV density), the situation is very different (see \cref{fig:matrix_models_tov}).
The Sly4, PCP(BSK24), DD2, SFHO, and BL models are incompatible with the higher $X_{\rm high}$-range. 
For these models, the sampled interpolations for the less restrictive $X_{\rm mid}$ and $X_{\rm low}$ ranges are significantly constrained.  

In the case of APR, the model EoS is terminated at a density below the maximum density reached in the TOV solution, specifically where the EoS first becomes acausal. 
This density corresponds to the central density of a NS with mass $M\approx2.07M_\odot$. 
Since the termination occurs at a lower density, the resulting extrapolations span a broader range.

In all of these cases (besides FSU2R) the behavior of the matter must have an abrupt change immediately the TOV density; the matter must undergo a dramatic softening resembling a strong first-order phase transition at densities immediately above those reached in NSs, which extends for a density interval $\Delta n \approx 15\, \ns$.
This is most clearly seen as the significant change of the slope of the trace anomaly. 
Note that since this transition takes place above the TOV density, it does not \emph{cause} the star to collapse; rather it is coincidental that the collapse and phase-transition-like behavior coincide, which seems highly unlikely. 
The change in thermodynamic behaviour is also reflected in the mass-radius plot, leading to a kink feature entering the unstable branch.  

We note that diffusing into the model region would lead to even more restrictive behaviour, as it would require a smooth connection to the stiff EoS, extending the high-$c_s^2$ behaviour of the model over a larger density range. 
In most cases, this would leave no viable extensions at all. Therefore, as mentioned before, allowing for abrupt behaviour at the TOV point is a more conservative way to treat the low-density limit.

In contrast, FSU2R permits extensions for the full range of $X$. 
This is a consequence of its relatively soft behaviour and positive trace anomaly all the way up to the TOV density. 
Unlike the other hadronic models considered here, however, the parameters of FSU2R were already informed by astrophysical observations.
It is important to note, that while this behaviour is compatible with pQCD constraints for larger values of $X$, the model predicts large radii above 12~km, which is in tension with the NICER radius measurements of PSR J0614$-$3329~\citep{Mauviard_2025} and PSR J0437$-$4715~\citep{Choudhury:2024xbk}.

The representative set of non-hadronic EoS models is shown in \cref{fig:exotica}, where each model is taken up to the TOV density. 
As evident from the figure, the behaviour required by the models at higher densities is generally less restricted than for purely hadronic models. 
However, BFH(QHC19-B), OPGR(DDHdeltaY4), and DS(CMF) remain sufficiently stiff that they require an abrupt change in the speed of sound at the TOV density, leading to effectively first-order-phase-transition-like extensions for the $X_{\rm high}$ range. 
In contrast, for the V-QCD and Quarkyonic models, the softening already occurs before the TOV point. 
As a result, the required behaviour above the TOV density is considerably less restrictive and does not force first-order-phase-transition-like behaviour.

\section{Conclusions}

In this study, we reveal what NS-matter models imply for the EoS behaviour beyond different termination densities by exploring the allowed range of possible valid extensions from the termination point to the perturbative-QCD limit. 
We find that for purely hadronic models, the behaviour remains largely unconstrained if the model is only trusted up to the central density of a $1.4M_\odot$ NS. 
However, if one assumes that hadronic models remain valid up to the maximal density reached in stable NSs, the situation changes dramatically. 
In this case, the EoS becomes incompatible with higher values of the pQCD renormalization-scale parameter $X$ and is forced into highly specific behaviour for allowed values of $X$.

This behaviour is characterized by an abrupt change at the TOV point, with an effectively first-order-phase-transition-like density jump of $\Delta n \approx 15\,\ns$. 
Importantly, this behaviour does not itself destabilize the star; rather, it emerges as a consequence of enforcing a causal, stable, and thermodynamically consistent connection to the pQCD limit.
Such a coincidental behaviour across all viable extensions seems highly unlikely. 
In contrast, this behaviour is not enforced if the EoS softens before the TOV density, as demonstrated by models exhibiting crossover or first-order phase transitions, such as V-QCD.

\begin{figure}
    \centering
    \includegraphics[width=1.0\linewidth]{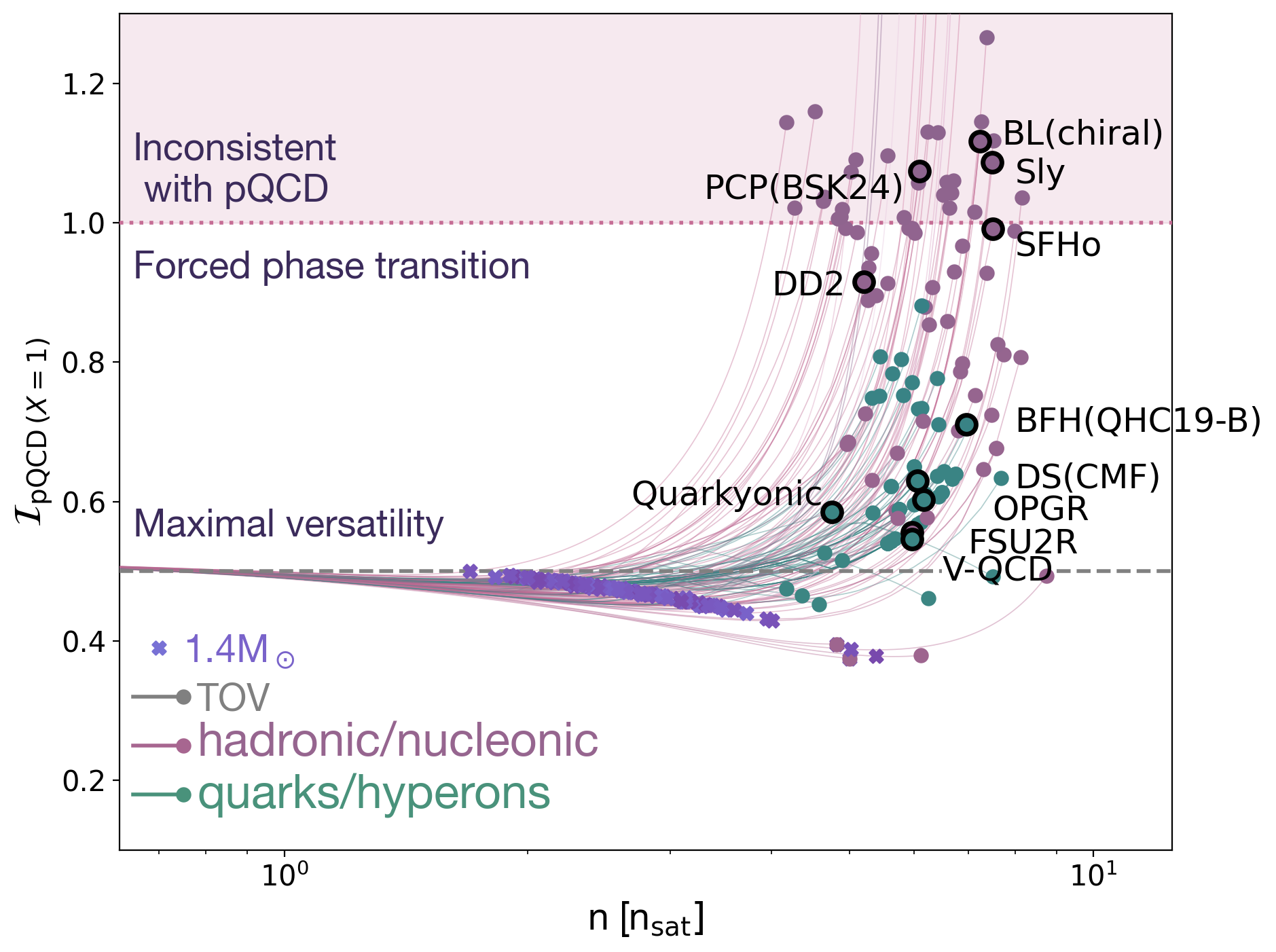}
    \caption{pQCD tension index as a function of central density for all zero-temperature, $\beta$-equilibrium EoSs available in the CompOSE database. Purple crosses indicate the central density of a $1.4M_\odot$ NS. Green dots correspond to the TOV point for EoSs that include tabulated quark or hyperonic degrees of freedom, while purple dots mark the TOV point of purely hadronic EoSs. Highlighted and labeled points correspond to the EoSs used in the previous figures.}
    \label{fig:ipqcd}
\end{figure}

While we investigated this behaviour only for a representative set of EoSs, the extremeness of the required extension to pQCD can be quantified for all zero-temperature, $\beta$-equilibrium EoSs available in the CompOSE database. 
This is shown in \cref{fig:ipqcd}, where the measure of extremeness is the pQCD tension index, defined in \cref{eq:iqcd}, shown as a function of central density. 
Values of $\ilh>1$ indicate that the model is inconsistent with pQCD for a given value of $X = 1$. 
Values of $\ilh=1$, or slightly below, correspond to the highly specific behaviour associated with a forced first-order-like transition.

The purple crosses in \cref{fig:ipqcd} indicate the central density of a $1.4M_\odot$ NS, where all models predict ${\ilh\approx0.5}$, corresponding to maximal versatility in the allowed behaviour. The green dots correspond to the TOV point for EoSs that include tabulated quark or hyperonic degrees of freedom, including the EoSs shown in \cref{fig:exotica}, which are additionally highlighted in the figure.

The purple dots correspond to the TOV point of the purely hadronic EoSs. 
Their clustering close to the $\ilh=1$ line suggests that the set of EoSs chosen for \cref{fig:matrix_models_tov} is representative of the broader trend: purely hadronic EoSs trusted up to the TOV point generically lead to large first-order-phase-transition-like behaviour above it. 
The exceptions to this trend, such as the highlighted FSU2R, are typically softer EoSs at intermediate densities. 

The EoSs showcased in this work have been widely used in binary NS-merger or core-collapse supernova simulations, as well as many other studies. 
Our results therefore imply that such simulations may carry an implicit assumption about the behaviour of QCD matter at higher densities: namely, that it undergoes a strong first-order-phase-transition-like change that happens to coincide with the TOV point. 
Since such behaviour appears highly artificial, avoiding it while remaining consistent with astrophysical constraints requires a softening of the EoS before the TOV point, for example through a crossover to quark matter~\citep{Annala:2023cwx, Fujimoto:2024ymt} or a first-order phase transition~\citep{Komoltsev:2024lcr,Blomqvist:2025cxe} at lower densities.

In conclusion, we have demonstrated that pQCD constraints provide useful guidance for NS-EoS model building. 
Our results indicate that models with additional degrees of freedom, leading to a softening of matter below $M_{\rm TOV}$, are preferred. 
Consequently, our work disfavors purely nucleonic matter and supports the appearance of additional degrees of freedom---without indicating a preference for any particular type---in stable massive NS.

\section*{Data Availability}

An interactive web application based on the framework developed in this work is available at \cite{GPBapp}. A Jupyter notebook implementation is also available on Zenodo \citep{gorda_2026_20797558}. The application generates a prior ensemble of allowed EoS extensions connecting a low-density neutron-star matter model to the high-density pQCD regime. 
Given a user-specified termination point, $(\mul, \nl, \pl)$, it constructs and visualizes the ensemble of allowed extensions.

\section*{Author Contributions}
Authors are listed in alphabetical order.

\begin{acknowledgments} 
    O.K.\ thanks Christian Ecker for the discussion and help. 
    O.K.\ acknowledges support from the Alexander von Humboldt Foundation through a Humboldt Research Fellowship for Postdoctoral Researchers.
    O.K.\ and J.S.B.\ acknowledge support by the Deutsche Forschungsgemeinschaft (DFG, German Research Foundation) through the 
    CRC-TR 211 'Strong-interaction matter under extreme conditions'– project number 315477589 – TRR 211. A.K. was supported by the Research Council of Norway through the FRIPRO programme (CoreQCD, project number 361873).
    This research was supported in part by grant NSF PHY-2309135 to the Kavli Institute for Theoretical Physics (KITP).
\end{acknowledgments}

\bibliographystyle{aasjournalv7}
\bibliography{references}

@article{Typel:2013rza,
    author = {Typel, S. and Oertel, M. and Kl{\"a}hn, T.},
    title = "{CompOSE CompStar online supernova equations of state harmonising the concert of nuclear physics and astrophysics compose.obspm.fr}",
    eprint = "1307.5715",
    archivePrefix = "arXiv",
    primaryClass = "astro-ph.SR",
    doi = "10.1134/S1063779615040061",
    journal = "Phys. Part. Nucl.",
    volume = "46",
    number = "4",
    pages = "633--664",
    year = "2015"
}

@article{Oertel:2016bki,
    author = {Oertel, M. and Hempel, M. and Kl{\"a}hn, T. and Typel, S.},
    title = "{Equations of state for supernovae and compact stars}",
    eprint = "1610.03361",
    archivePrefix = "arXiv",
    primaryClass = "astro-ph.HE",
    doi = "10.1103/RevModPhys.89.015007",
    journal = "Rev. Mod. Phys.",
    volume = "89",
    number = "1",
    pages = "015007",
    year = "2017"
}

@article{CompOSECoreTeam:2022ddl,
    author = "Typel, S. and others",
    collaboration = "CompOSE Core Team",
    title = "{CompOSE Reference Manual}",
    eprint = "2203.03209",
    archivePrefix = "arXiv",
    primaryClass = "astro-ph.HE",
    doi = "10.1140/epja/s10050-022-00847-y",
    journal = "Eur. Phys. J. A",
    volume = "58",
    number = "11",
    pages = "221",
    year = "2022"
}

@article{deForcrand:2009zkb,
    author = "de Forcrand, Philippe",
    editor = "Liu, Chuan and Zhu, Yu",
    title = "{Simulating QCD at finite density}",
    eprint = "1005.0539",
    archivePrefix = "arXiv",
    primaryClass = "hep-lat",
    reportNumber = "CERN-PH-TH-2010-090",
    doi = "10.22323/1.091.0010",
    journal = "PoS",
    volume = "LAT2009",
    pages = "010",
    year = "2009"
}

@article{Gorda:2023mkk,
    author = {Gorda, Tyler and Paatelainen, Risto and S{\"a}ppi, Saga and Sepp{\"a}nen, Kaapo},
    title = "{Equation of State of Cold Quark Matter to $O(\alpha_s^3 \ln \alpha_s)$}",
    eprint = "2307.08734",
    archivePrefix = "arXiv",
    primaryClass = "hep-ph",
    reportNumber = "HIP-2023-10/TH, TUM-EFT 181/23",
    doi = "10.1103/PhysRevLett.131.181902",
    journal = "Phys. Rev. Lett.",
    volume = "131",
    number = "18",
    pages = "181902",
    year = "2023"
}

@article{Semposki:2025etb,
    author = "Semposki, A. C. and Drischler, C. and Furnstahl, R. J. and Phillips, D. R.",
    title = "{Microscopic constraints for the equation~of state and structure of neutron stars: A Bayesian model mixing framework}",
    eprint = "2505.18921",
    archivePrefix = "arXiv",
    primaryClass = "nucl-th",
    doi = "10.1103/fxv6-gdnw",
    journal = "Phys. Rev. C",
    volume = "113",
    number = "1",
    pages = "015808",
    year = "2026"
}

@article{Gorda:2022jvk,
    author = "Gorda, Tyler and Komoltsev, Oleg and Kurkela, Aleksi",
    title = "{Ab-initio QCD Calculations Impact the Inference of the Neutron-star-matter Equation of State}",
    eprint = "2204.11877",
    archivePrefix = "arXiv",
    primaryClass = "nucl-th",
    doi = "10.3847/1538-4357/acce3a",
    journal = "Astrophys. J.",
    volume = "950",
    number = "2",
    pages = "107",
    year = "2023"
}

@article{Komoltsev:2021jzg,
    author = "Komoltsev, Oleg and Kurkela, Aleksi",
    title = "{How Perturbative QCD Constrains the Equation of State at Neutron-Star Densities}",
    eprint = "2111.05350",
    archivePrefix = "arXiv",
    primaryClass = "nucl-th",
    doi = "10.1103/PhysRevLett.128.202701",
    journal = "Phys. Rev. Lett.",
    volume = "128",
    number = "20",
    pages = "202701",
    year = "2022"
}

@article{Gorda:2021znl,
    author = {Gorda, Tyler and Kurkela, Aleksi and Paatelainen, Risto and S{\"a}ppi, Saga and Vuorinen, Aleksi},
    title = "{Soft Interactions in Cold Quark Matter}",
    eprint = "2103.05658",
    archivePrefix = "arXiv",
    primaryClass = "hep-ph",
    reportNumber = "HIP-2021-9/TH",
    doi = "10.1103/PhysRevLett.127.162003",
    journal = "Phys. Rev. Lett.",
    volume = "127",
    number = "16",
    pages = "162003",
    year = "2021"
}

@article{Gorda:2023usm,
    author = "Gorda, Tyler and Komoltsev, Oleg and Kurkela, Aleksi and Mazeliauskas, Aleksas",
    title = "{Bayesian uncertainty quantification of perturbative QCD input to the neutron-star equation of state}",
    eprint = "2303.02175",
    archivePrefix = "arXiv",
    primaryClass = "hep-ph",
    doi = "10.1007/JHEP06(2023)002",
    journal = "JHEP",
    volume = "06",
    pages = "002",
    year = "2023"
}

@article{Komoltsev:2023zor,
    author = "Komoltsev, Oleg and Somasundaram, Rahul and Gorda, Tyler and Kurkela, Aleksi and Margueron, J{\'e}r{\^o}me and Tews, Ingo",
    title = "{Equation of state at neutron-star densities and beyond from perturbative QCD}",
    eprint = "2312.14127",
    archivePrefix = "arXiv",
    primaryClass = "nucl-th",
    reportNumber = "LA-UR-23-32625",
    doi = "10.1103/PhysRevD.109.094030",
    journal = "Phys. Rev. D",
    volume = "109",
    number = "9",
    pages = "094030",
    year = "2024"
}

@article{Annala:2019puf,
    author = {Annala, Eemeli and Gorda, Tyler and Kurkela, Aleksi and N{\"a}ttil{\"a}, Joonas and Vuorinen, Aleksi},
    title = "{Evidence for quark-matter cores in massive neutron stars}",
    eprint = "1903.09121",
    archivePrefix = "arXiv",
    primaryClass = "astro-ph.HE",
    reportNumber = "CERN-TH-2019-031, HIP-2019-7/TH",
    doi = "10.1038/s41567-020-0914-9",
    journal = "Nature Phys.",
    volume = "16",
    number = "9",
    pages = "907--910",
    year = "2020"
}

@article{Annala:2023cwx,
    author = {Annala, Eemeli and Gorda, Tyler and Hirvonen, Joonas and Komoltsev, Oleg and Kurkela, Aleksi and N{\"a}ttil{\"a}, Joonas and Vuorinen, Aleksi},
    title = "{Strongly interacting matter exhibits deconfined behavior in massive neutron stars}",
    eprint = "2303.11356",
    archivePrefix = "arXiv",
    primaryClass = "astro-ph.HE",
    reportNumber = "HIP-2023-5/TH, HIP-2023-5/TH",
    doi = "10.1038/s41467-023-44051-y",
    journal = "Nature Commun.",
    volume = "14",
    number = "1",
    pages = "8451",
    year = "2023"
}

@article{Choudhury:2024xbk,
    author = "Choudhury, Devarshi and Salmi, Tuomo and Vinciguerra, Serena and others",
    title = "{A NICER View of the Nearest and Brightest Millisecond Pulsar: PSR J0437{\textendash}4715}",
    eprint = "2407.06789",
    archivePrefix = "arXiv",
    primaryClass = "astro-ph.HE",
    doi = "10.3847/2041-8213/ad5a6f",
    journal = "Astrophys. J. Lett.",
    volume = "971",
    number = "1",
    pages = "L20",
    year = "2024"
}

@article{Komoltsev:2024lcr,
    author = "Komoltsev, Oleg",
    title = "{First-order phase transitions in the cores of neutron stars}",
    eprint = "2404.05637",
    archivePrefix = "arXiv",
    primaryClass = "nucl-th",
    doi = "10.1103/PhysRevD.110.L071502",
    journal = "Phys. Rev. D",
    volume = "110",
    number = "7",
    pages = "L071502",
    year = "2024"
}

@article{Gorda:2025aiu,
    author = "Gorda, Tyler and Komoltsev, Oleg and Kurkela, Aleksi and Sunde, Eirik",
    title = "{Constrained Gaussian-process-bridge Prior for Neutron-star Equation-of-state Inference}",
    eprint = "2512.18044",
    archivePrefix = "arXiv",
    primaryClass = "astro-ph.HE",
    doi = "10.3847/1538-4357/ae552a",
    journal = "Astrophys. J.",
    volume = "1002",
    number = "1",
    pages = "40",
    year = "2026"
}

@article{Semposki:2024vnp,
    author = "Semposki, A. C. and Drischler, C. and Furnstahl, R. J. and Melendez, J. A. and Phillips, D. R.",
    title = "{From chiral effective field theory to perturbative QCD: A Bayesian model mixing approach to symmetric nuclear matter}",
    eprint = "2404.06323",
    archivePrefix = "arXiv",
    primaryClass = "nucl-th",
    doi = "10.1103/PhysRevC.111.035804",
    journal = "Phys. Rev. C",
    volume = "111",
    number = "3",
    pages = "035804",
    year = "2025"
}

@article{Antonopoulou:2022yot,
    author = "Antonopoulou, Danai and Bozzo, Enrico and Ishizuka, Chikako and Jones, David Ian and Oertel, Micaela and Providencia, Constan{\c{c}}a and Tolos, Laura and Typel, Stefan",
    title = "{CompOSE: a repository for neutron star equations of state and transport properties}",
    doi = "10.1140/epja/s10050-022-00908-2",
    journal = "Eur. Phys. J. A",
    volume = "58",
    number = "12",
    pages = "254",
    year = "2022"
}

@misc{MUSES_web,
    key = "MUSES Project Website",
    url = "https://musesframework.io/",
    year = "2024"
}

@article{MUSES:2023hyz,
    author = "Kumar, Rajesh and others",
    collaboration = "MUSES",
    title = "{Theoretical and experimental constraints for the equation of state of dense and hot matter}",
    eprint = "2303.17021",
    archivePrefix = "arXiv",
    primaryClass = "nucl-th",
    doi = "10.1007/s41114-024-00049-6",
    journal = "Living Rev. Rel.",
    volume = "27",
    number = "1",
    pages = "3",
    year = "2024"
}

@article{Han:2022rug,
    author = "Han, Ming-Zhe and Huang, Yong-Jia and Tang, Shao-Peng and Fan, Yi-Zhong",
    title = "{Plausible presence of new state in neutron stars with masses above 0.98MTOV}",
    eprint = "2207.13613",
    archivePrefix = "arXiv",
    primaryClass = "astro-ph.HE",
    reportNumber = "RIKEN-iTHEMS-Report-23",
    doi = "10.1016/j.scib.2023.04.007",
    journal = "Sci. Bull.",
    volume = "68",
    pages = "913--919",
    year = "2023"
}

@article{Akmal:1998cf,
    author = "Akmal, A. and Pandharipande, V. R. and Ravenhall, D. G.",
    title = "{The Equation of state of nucleon matter and neutron star structure}",
    eprint = "nucl-th/9804027",
    archivePrefix = "arXiv",
    doi = "10.1103/PhysRevC.58.1804",
    journal = "Phys. Rev. C",
    volume = "58",
    pages = "1804--1828",
    year = "1998"
}

@article{Bombaci:2018ksa,
    author = "Bombaci, Ignazio and Logoteta, Domenico",
    title = "{Equation of state of dense nuclear matter and neutron star structure from nuclear chiral interactions}",
    eprint = "1805.11846",
    archivePrefix = "arXiv",
    primaryClass = "astro-ph.HE",
    doi = "10.1051/0004-6361/201731604",
    journal = "Astron. Astrophys.",
    volume = "609",
    pages = "A128",
    year = "2018"
}

@article{Pearson:2018tkr,
    author = "Pearson, J. M. and Chamel, N. and Potekhin, A. Y. and Fantina, A. F. and Ducoin, C. and Dutta, A. K. and Goriely, S.",
    title = "{Unified equations of state for cold non-accreting neutron stars with Brussels{\textendash}Montreal functionals {\textendash} I. Role of symmetry energy}",
    eprint = "1903.04981",
    archivePrefix = "arXiv",
    primaryClass = "astro-ph.HE",
    doi = "10.1093/mnras/sty2413",
    journal = "Mon. Not. Roy. Astron. Soc.",
    volume = "481",
    number = "3",
    pages = "2994--3026",
    year = "2018",
    note = "[Erratum: Mon.Not.Roy.Astron.Soc. 486, 768 (2019)]"
}

@article{Gulminelli:2015csa,
    author = "Gulminelli, F. and Raduta, Ad. R.",
    title = "{Unified treatment of subsaturation stellar matter at zero and finite temperature}",
    eprint = "1504.04493",
    archivePrefix = "arXiv",
    primaryClass = "nucl-th",
    doi = "10.1103/PhysRevC.92.055803",
    journal = "Phys. Rev. C",
    volume = "92",
    number = "5",
    pages = "055803",
    year = "2015"
}

@article{Chabanat:1997un,
    author = "Chabanat, E. and Bonche, P. and Haensel, P. and Meyer, J. and Schaeffer, R.",
    title = "{A Skyrme parametrization from subnuclear to neutron star densities. 2. Nuclei far from stablities}",
    doi = "10.1016/S0375-9474(98)00180-8",
    journal = "Nucl. Phys. A",
    volume = "635",
    pages = "231--256",
    year = "1998",
    note = "[Erratum: Nucl.Phys.A 643, 441--441 (1998)]"
}

@article{HEMPEL2010210,
title = {A statistical model for a complete supernova equation of state},
journal = {Nuclear Physics A},
volume = {837},
number = {3},
pages = {210-254},
year = {2010},
issn = {0375-9474},
doi = {https://doi.org/10.1016/j.nuclphysa.2010.02.010},
url = {https://www.sciencedirect.com/science/article/pii/S0375947410003325},
author = {Matthias Hempel and Jürgen Schaffner-Bielich}
}

@article{Steiner:2012rk,
    author = "Steiner, Andrew W. and Hempel, Matthias and Fischer, Tobias",
    title = "{Core-collapse supernova equations of state based on neutron star observations}",
    eprint = "1207.2184",
    archivePrefix = "arXiv",
    primaryClass = "astro-ph.SR",
    reportNumber = "INT-PUB-12-033",
    doi = "10.1088/0004-637X/774/1/17",
    journal = "Astrophys. J.",
    volume = "774",
    pages = "17",
    year = "2013"
}

@article{Negreiros_2018,
doi = {10.3847/1538-4357/aad049},
url = {https://doi.org/10.3847/1538-4357/aad049},
year = {2018},
month = {aug},
publisher = {The American Astronomical Society},
volume = {863},
number = {1},
pages = {104},
author = {Negreiros, Rodrigo and Tolos, Laura and Centelles, Mario and Ramos, Angels and Dexheimer, Veronica},
title = {Cooling of Small and Massive Hyperonic Stars},
journal = {The Astrophysical Journal}
}

@article{Baym_2019,
doi = {10.3847/1538-4357/ab441e},
url = {https://doi.org/10.3847/1538-4357/ab441e},
year = {2019},
month = {oct},
publisher = {The American Astronomical Society},
volume = {885},
number = {1},
pages = {42},
author = {Baym, Gordon and Furusawa, Shun and Hatsuda, Tetsuo and Kojo, Toru and Togashi, Hajime},
title = {New Neutron Star Equation of State with Quark–Hadron Crossover},
journal = {The Astrophysical Journal}
}

@article{Oertel_2015,
doi = {10.1088/0954-3899/42/7/075202},
url = {https://doi.org/10.1088/0954-3899/42/7/075202},
year = {2015},
month = {jun},
publisher = {IOP Publishing},
volume = {42},
number = {7},
pages = {075202},
author = {Oertel, M and Providência, C and Gulminelli, F and Raduta, Ad R},
title = {Hyperons in neutron star matter within relativistic mean-field models},
journal = {Journal of Physics G: Nuclear and Particle Physics}
}

@article{PhysRevX.12.041012,
  title = {Dense and Hot QCD at Strong Coupling},
  author = {Demircik, Tuna and Ecker, Christian and J\"arvinen, Matti},
  journal = {Phys. Rev. X},
  volume = {12},
  issue = {4},
  pages = {041012},
  numpages = {22},
  year = {2022},
  month = {Oct},
  publisher = {American Physical Society},
  doi = {10.1103/PhysRevX.12.041012},
  url = {https://link.aps.org/doi/10.1103/PhysRevX.12.041012}
}

@article{PhysRevC.81.045201,
  title = {Novel approach to modeling hybrid stars},
  author = {Dexheimer, V. A. and Schramm, S.},
  journal = {Phys. Rev. C},
  volume = {81},
  issue = {4},
  pages = {045201},
  numpages = {5},
  year = {2010},
  month = {Apr},
  publisher = {American Physical Society},
  doi = {10.1103/PhysRevC.81.045201},
  url = {https://link.aps.org/doi/10.1103/PhysRevC.81.045201}
}

@article{PhysRevC.103.025808,
  title = {GW190814 as a massive rapidly rotating neutron star with exotic degrees of freedom},
  author = {Dexheimer, V. and Gomes, R. O. and Kl\"ahn, T. and Han, S. and Salinas, M.},
  journal = {Phys. Rev. C},
  volume = {103},
  issue = {2},
  pages = {025808},
  numpages = {10},
  year = {2021},
  month = {Feb},
  publisher = {American Physical Society},
  doi = {10.1103/PhysRevC.103.025808},
  url = {https://link.aps.org/doi/10.1103/PhysRevC.103.025808}
}

@article{Mauviard_2025,
doi = {10.3847/1538-4357/ae145d},
url = {https://doi.org/10.3847/1538-4357/ae145d},
year = {2025},
month = {dec},
publisher = {The American Astronomical Society},
volume = {995},
number = {1},
pages = {60},
author = {Mauviard, Lucien and Guillot, Sebastien and Salmi, Tuomo and Choudhury, Devarshi and Dorsman, Bas and González-Caniulef, Denis and Hoogkamer, Mariska and Huppenkothen, Daniela and Kazantsev, Christine and Kini, Yves and Olive, Jean-Francois and Stammler, Pierre and Watts, Anna L. and Mendes, Melissa and Rutherford, Nathan and Schwenk, Achim and Svensson, Isak and Bogdanov, Slavko and Kerr, Matthew and Ray, Paul S. and Guillemot, Lucas and Cognard, Ismaël and Theureau, Gilles},
title = {A NICER View of the 1.4 $M\_odot$ Edge-on Pulsar PSR J0614-3329},
journal = {The Astrophysical Journal}
}

@article{PhysRevD.102.023021,
  title = {Quarkyonic matter equation of state in beta-equilibrium},
  author = {Zhao, Tianqi and Lattimer, James M.},
  journal = {Phys. Rev. D},
  volume = {102},
  issue = {2},
  pages = {023021},
  numpages = {12},
  year = {2020},
  month = {Jul},
  publisher = {American Physical Society},
  doi = {10.1103/PhysRevD.102.023021},
  url = {https://link.aps.org/doi/10.1103/PhysRevD.102.023021}
}

@misc{zhao_quarkyonic_code,
  author       = {Tianqi Zhao},
  title        = {quaryonic\_eos: n-u-d version of quarkyonic matter EOS},
  year         = {2020},
  howpublished = {\url{https://github.com/sotzee/quaryonic_eos}},
  note         = {GitHub repository, accessed 2026-06-10}
}

@misc{GPBapp,
  author       = {Gorda, Tyler and Komoltsev, Oleg and Kurkela, Aleksi and Schaffner-Bielich, Jürgen},
  title        = {Gaussian-Process Bridge Extension for Neutron-Star Equations of State},
  year         = {2026},
  howpublished = {\url{https://gaussian-process-bridge-eos-sampler.streamlit.app/}},
  note         = {Interactive web application}
}

@article{Nagata:2021ugx,
    author = "Nagata, Keitaro",
    title = "{Finite-density lattice QCD and sign problem: Current status and open problems}",
    eprint = "2108.12423",
    archivePrefix = "arXiv",
    primaryClass = "hep-lat",
    doi = "10.1016/j.ppnp.2022.103991",
    journal = "Prog. Part. Nucl. Phys.",
    volume = "127",
    pages = "103991",
    year = "2022"
}

@unpublished{Blomqvist:2025cxe,
    author = "Blomqvist, Sofia and Ecker, Christian and Gorda, Tyler and Vuorinen, Aleksi",
    title = "{Strong model-agnostic constraints for twin-star solutions}",
    eprint = "2512.19477",
    archivePrefix = "arXiv",
    primaryClass = "astro-ph.HE",
    month = "12",
    year = "2025"
}

@article{PhysRevC.90.044305,
  title = {Building relativistic mean field models for finite nuclei and neutron stars},
  author = {Chen, Wei-Chia and Piekarewicz, J.},
  journal = {Phys. Rev. C},
  volume = {90},
  issue = {4},
  pages = {044305},
  numpages = {17},
  year = {2014},
  month = {Oct},
  publisher = {American Physical Society},
  doi = {10.1103/PhysRevC.90.044305},
  url = {https://link.aps.org/doi/10.1103/PhysRevC.90.044305}
}

@article{Clevinger:2022xzl,
    author = "Clevinger, A. and Corkish, J. and Aryal, K. and Dexheimer, V.",
    title = "{Hybrid equations of state for neutron stars with hyperons and deltas}",
    eprint = "2205.00559",
    archivePrefix = "arXiv",
    primaryClass = "astro-ph.HE",
    doi = "10.1140/epja/s10050-022-00745-3",
    journal = "Eur. Phys. J. A",
    volume = "58",
    number = "5",
    pages = "96",
    year = "2022"
}

@article{Typel:2009sy,
    author = "Typel, S. and Ropke, G. and Klahn, T. and Blaschke, D. and Wolter, H. H.",
    title = "{Composition and thermodynamics of nuclear matter with light clusters}",
    eprint = "0908.2344",
    archivePrefix = "arXiv",
    primaryClass = "nucl-th",
    doi = "10.1103/PhysRevC.81.015803",
    journal = "Phys. Rev. C",
    volume = "81",
    pages = "015803",
    year = "2010"
}

@misc{gorda_2026_20797558,
  author       = {Gorda, Tyler and
                  Komoltsev, Oleg and
                  Kurkela, Aleksi and
                  Schaffner-Bielich, Juergen},
  title        = {Gaussian-Process Bridge Extension for Neutron-Star
                   Equations of State
                  },
  month        = jun,
  year         = 2026,
  publisher    = {Zenodo},
  doi          = {10.5281/zenodo.20797558},
  url          = {https://doi.org/10.5281/zenodo.20797558},
}

@article{Fujimoto:2024ymt,
    author = "Fujimoto, Yuki and Fukushima, Kenji and Hotokezaka, Kenta and Kyutoku, Koutarou",
    title = "{Signature of hadron-quark crossover in binary-neutron-star mergers}",
    eprint = "2408.10298",
    archivePrefix = "arXiv",
    primaryClass = "astro-ph.HE",
    reportNumber = "INT-PUB-24-041",
    doi = "10.1103/PhysRevD.111.063054",
    journal = "Phys. Rev. D",
    volume = "111",
    number = "6",
    pages = "063054",
    year = "2025"
}

\end{document}